\documentclass{PoS}
\usepackage{amsmath,amssymb,graphicx,slashed,wrapfig}
\title{The Static Approximation to B Meson Mixing \\ using Light
Domain-Wall Fermions: Perturbative Renormalization and Ground State
Degeneracies}

\ShortTitle{The Static Approximation to B Meson Mixing using Light
DWF}

\author{Norman H.~Christ, \speaker{Thomas T.~Dumitrescu}, Oleg Loktik\\
        Department of Physics   , Columbia University, New York, NY 10027, USA\\
        E-mail: \email{nhc@phys.columbia.edu}, \email{td2135@columbia.edu}, \email{oleg@phys.columbia.edu}}

\author{Taku Izubuchi \\
        RIKEN-BNL Research Center, Brookhaven National Laboratory,
        Upton, NY 11973, USA and Institute for Theoretical Physics,
        Kanazawa University, Kanazawa, Ishikawa 920-1192, Japan \\
        E-mail: \email{izubuchi@hep.s.kanazawa-u.ac.jp}}

\author{RBC-UKQCD Collaboration}

\abstract{We discuss the theoretical input into the current
RBC-UKQCD calculation of $f_{B_{d,\,s}}$ and $B_{B_{d,\,s}}$ using a
smeared static heavy quark propagator, light domain-wall quarks and
the Iwasaki gauge action. We present the complete one-loop,
mean-field improved matching of heavy-light current and four-fermion
lattice operators onto the static continuum theory renormalized in
$\overline{\text{MS}}$(NDR). The large degeneracies present in a
static calculation are addressed, and a method for extracting $f_B$
and $B_B$ using only box sources is described; implications for
future calculations are discussed.}

\FullConference{The XXV International Symposium on Lattice Field Theory\\
         July 30-4 August 2007\\
         Regensburg, Germany}

\begin{document}

\section{Introduction}

Precision measurements of the CKM matrix put the Standard Model to a
stringent test and constrain possible physics beyond it. Using the
measured frequency of $B_{q} - \overline{B}_{q}, \; q \in \{d,s\}$
oscillations to determine the CKM matrix elements $|V_{tq}|$
requires a reliable lattice calculation of the non-perturbative $B_q
- \overline{B}_q$ mixing matrix elements $\frac{8}{3}m^2_{B_q}
f^{\,2}_{B_q} B_{B_q}$. A $2+1$ flavor, unquenched calculation of
$f_{B_q}$ and $B_{B_q}$ has been carried out by the RBC-UKQCD
collaboration in the infinite heavy quark mass limit using light
domain-wall fermions on a $(2 \; \rm{fm})^3$ spatial volume
\cite{Wennekers:2007,AokiRBC:2007}; this is currently being extended
to a $(3 \; \rm{fm})^3$ spatial volume and towards physical light
quark masses \cite{Aoki:2007}. In the following, we discuss the
perturbative lattice-continuum matching of the operators relevant
for the RBC-UKQCD calculation, following in part the detailed
discussion in Refs.~\cite{AokiRBC:2007,Loktik:2006kz}. We also point
out the subtle degeneracy of heavy-light meson ground states, and
discuss its implications for the extraction of $f_B$ and $B_B$ from
lattice correlation functions.

\section{Action and Feynman Rules}
\label{sec:action}

\paragraph{}

The heavy $b$ quark is described by an improved lattice version of
the static limit of heavy quark effective theory with smeared,
SU(3)-projected gauge links $\overline{V}_0(\vec{x},t)$ to reduce
noise:

\begin{equation}\label{eq:hqa_imp}
S_{\text{static}} = \sum_{\vec{x}, \, t} \; \overline{h}(\vec{x},t +
a) \left[ h(\vec{x},t + a) -
\overline{V}^{\dagger}_0(\vec{x},t)h(\vec{x},t) \right].
\end{equation}

%
%
%

\noindent The SU(3) projection (discussed in
Ref.~\cite{AokiRBC:2007}) simplifies perturbative calculations by
allowing the smeared gauge links to be expanded in terms of an
effective gauge field $B^{\,a}_0(\vec{x}, t)$; in momentum space
$B^{\,a}_0(q) = h_\mu(q) A^{\,a}_{\mu}(q)$, where $A^{\,a}_{\mu}(q)$
is the physical gauge field and $h_\mu(q)$ is a form factor
depending on the smearing scheme. We focus on one of the two schemes
used in the RBC-UKQCD calculation (one-level APE blocking with
parameter $\alpha = 1$), resulting in a heavy quark gluon vertex

\begin{equation}
Y^a_{\mu}(k,k') = -ig_0T^a \delta_{\mu0} e^{-i(k_0+k_0')/2}
\rightarrow \overline{Y}^{\,a}_{\mu}(k,k') = -ig_0T^a h_{\mu}(q)
e^{-i(k_0+k_0')/2},
\end{equation}

\noindent where $g_0$ is the bare lattice coupling, $q$ is the gluon
momentum, and $h_{\mu}(q)$ is given by

\begin{equation}
\label{eq:formfactor} h_{\mu}(q) = (h_0(q), \, h_j(q)) = \left(1 -
\frac{2}{3}\sum_{l = 1}^3 \sin^{2}\left(\frac{q_l}{2}\right), \,
\frac{2}{3}\sin\left(\frac{q_0}{2}\right)
\sin\left(\frac{q_j}{2}\right)\right).
\end{equation}

\noindent The heavy quark two-gluon vertex and the heavy quark
propagator are given in Ref.~\cite{Loktik:2006kz}.


\paragraph{}

The light quarks are described by the domain-wall fermion action.
Each light flavor is represented by a $(4+1)$-dimensional
Wilson-style fermion field $\psi_s(\vec{x}, t)$ where $1 \leq s \leq
N$ labels the coordinate in the fifth dimension. The physical quark
field $q(\vec{x}, t)$ is constructed from chiral surface states at
$s = 1$ and $s = N$ via $q(\vec{x}, t) = P_R \psi_1(\vec{x}, t) +
P_{L} \psi_N(\vec{x}, t)$. The domain-wall height $M_5$ is a fixed
parameter of the theory; we set $M_5 = 1.8$ to match the RBC-UKQCD
calculation. A detailed description of domain-wall fermions and
their perturbative treatment for our choice of gauge action is given
in Ref.~\cite{Loktik:2006kz} and references therein, especially
Ref.~\cite{Aoki:2002iq}. In the perturbative calculation the light
quark masses were set to zero and the size $N$ of the fifth
dimension was taken to be large, resulting in an exact chiral
symmetry as $N \rightarrow \infty$. The gluons were described by the
Iwasaki gauge action, whose Feynman rules are given in
Ref.~\cite{Loktik:2006kz}.

%
%

\section{Perturbative Lattice-Continuum Matching at One-Loop}

The full QCD operators relevant for the extraction of $f_B$ and
$B_B$, defined in $\overline{\text{MS}}$(NDR) at the scale $\mu_b =
m_b$ of the $b$ quark mass, are the axial vector current $A_\rho =
\overline b \gamma_\rho \gamma_5 q$ and the parity-even part of the
$\Delta B = 2$ vector-axial four-quark operator:

\begin{equation}
 \left[\,\overline b \gamma^{\,\rho}(1-\gamma_5)q\right]
 \left[\,\overline b \gamma_\rho(1-\gamma_5)q\right] \rightarrow O_{VV+AA} =   
  \left(\overline b\gamma^{\,\rho} q\right)\left(\overline b \gamma_\rho q\right) +
  \left(\overline b\gamma^{\,\rho}\gamma_5q\right)
  \left(\overline b\gamma_\rho\gamma_5q\right).
\end{equation}

\noindent We match these operators at the scale $\mu_b$ to lattice
operators in the static effective theory (described in
Sec.~\ref{sec:action}) at the lattice scale $a^{-1}$ via the
continuum version of the static effective theory renormalized at a
scale $\mu$. Throughout our one-loop calculation we choose to set
$\mu = a^{-1}$; in the RBC-UKQCD calculation, the lattice scale is
given by $a^{-1} = 1.62$ GeV. The full QCD operators are related to
continuum static operators by

\begin{equation}
\label{eq:full_A_matching} A_\rho(\mu_b) =
 C_A(\mu_b,\mu) \widetilde A_\rho(\mu) +
 \mathcal O(\Lambda_\mathrm{QCD}/\mu_b),
\end{equation}
\begin{equation}
\label{eq:full_vv+aa_matching} O_{VV+AA}(\mu_b) =
  Z_1(\mu_b,\mu) \widetilde{O}_{VV+AA}(\mu) +
  Z_2(\mu_b,\mu) \widetilde{O}_{SS+PP}(\mu) +
  \mathcal O(\Lambda_\mathrm{QCD}/\mu_b).
\end{equation}

\noindent In terms of the static quark and antiquark fields
$h^{(\pm)}(x) = e^{\pm i m_b v\cdot x}(1 \pm \slashed{v}) b(x) / 2$
and for $m_b \rightarrow \infty$,

\begin{equation}
\widetilde A_\rho = \overline h^{(+)} \gamma_\rho\gamma_5 q,
\end{equation}
\begin{equation}
\label{eq:statOVVAA} \widetilde O_{VV+AA} =
 2\left(\overline h^{(+)}\gamma^{\,\rho} q\right)
  \left(\overline h^{(-)}\gamma_\rho q\right) +
 2\left(\overline h^{(+)}\gamma^{\,\rho}\gamma_5q\right)
  \left(\overline h^{(-)}\gamma_\rho\gamma_5q\right),
\end{equation}
\begin{equation}
\label{eq:statOSSPP} \widetilde O_{SS+PP} =
 2\left(\overline h^{(+)}q \right)\left(\overline h^{(-)}q\right) +
 2\left(\overline h^{(+)}\gamma_5q\right)\left(\overline
 h^{(-)}\gamma_5q\right).
\end{equation}

\noindent The static effective action discussed in
Sec.~\ref{sec:action} describes $h^{(+)}$ with $v = (1, \vec{0})$,
corresponding to a stationary meson. 
The constants $C_A(\mu_b,\mu)$ and $Z_{1,2}(\mu_b,\mu)$ are known at
one-loop; they are summarized in Ref.~\cite{AokiRBC:2007}. Using the
latest PDG values for $\alpha^{\overline{\text{MS}}}_s(m_Z)$ and
$m_b$, and running the coupling down at four-loops with the physical
number of flavors to determine
$\alpha^{\overline{\text{MS}}}_s(\mu_b)$ and
$\alpha^{\overline{\text{MS}}}_s(\mu)$ we obtain $C_A =  1.057, \;
Z_1 = 0.934 , \; Z_2 = -0.151$.

\paragraph{}

\begin{wrapfigure}{r}{6.0cm}
\begin{center}
\includegraphics{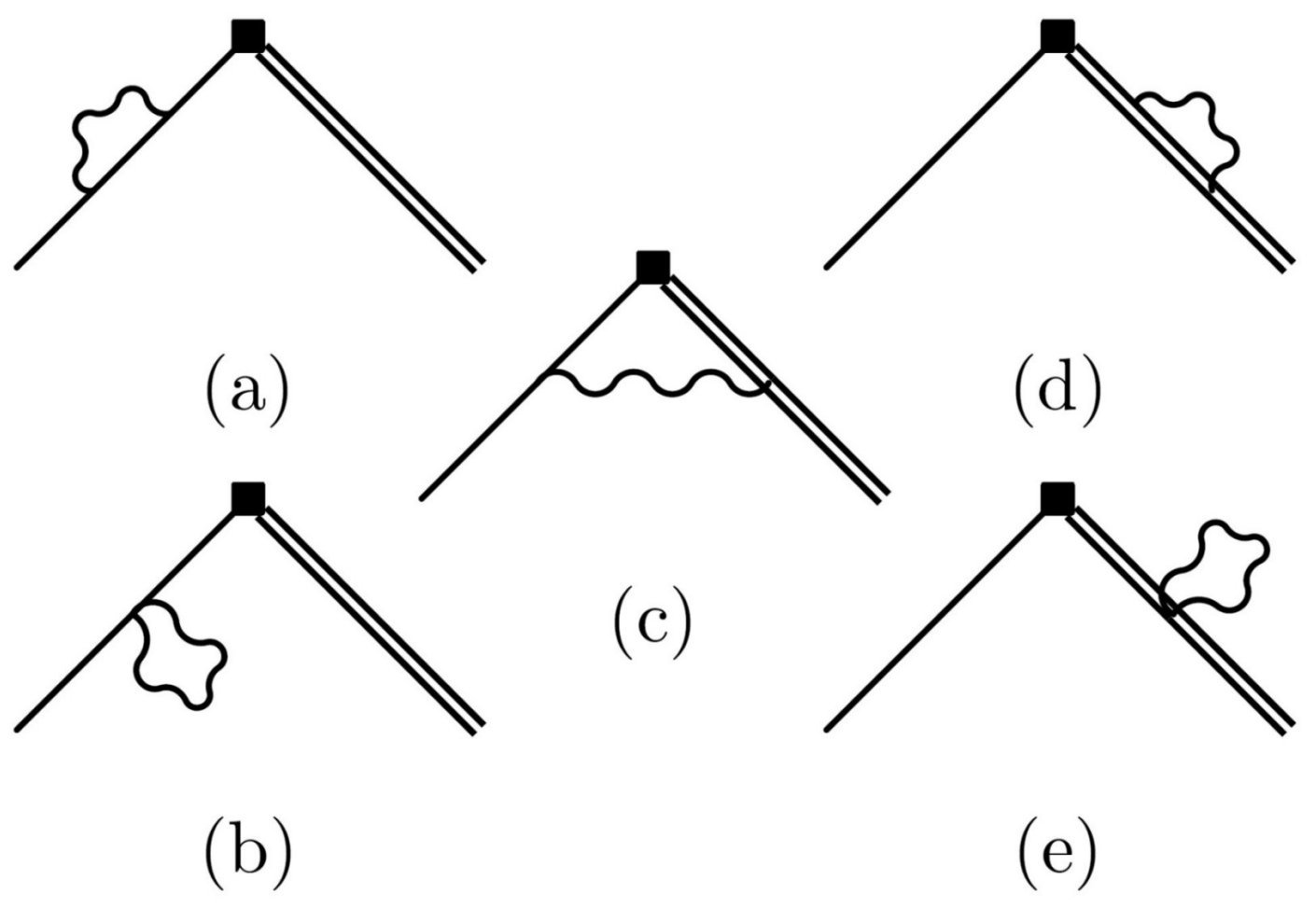}
\end{center}
\caption{\label{fig:heavylight} One-Loop Corrections to the
Heavy-Light Axial Current}
\end{wrapfigure}

We now describe the matching $\widetilde A_\rho(\mu) =
\widetilde{C}_{A}(\mu,a^{-1}) a^{-3}A^{\text{lat}}_\rho$ of the
heavy-light axial currents $\widetilde A_\rho(\mu)$ and
$A^{\text{lat}}_\rho$ (which is dimensionless) in the continuum and
lattice versions of the static effective theory. Results for the
four-fermion operators are summarized at the end of this section. We
compare the correlation function $\langle (\overline{h}(x)\Gamma
q(x))h(y)\overline{q}(z)\rangle$ in both theories; in this
discussion only one heavy quark field $h^{(+)} \equiv h$ enters. For
the axial current $\Gamma = \gamma_\rho\gamma_5$, but the light
quark chiral symmetry and the heavy quark spin symmetry $h
\rightarrow e^{-i\phi_j \epsilon_{jkl} \sigma_{kl}}h$ of both the
continuum and the lattice theory render the matching
$\Gamma$-independent. At one-loop and for small external quark
momenta $p \simeq 0$ the continuum and lattice correlation functions
are

\begin{equation}
\langle(\overline{h}\Gamma q)h\overline{q}\rangle = \frac{Z_h}{i
p_0}\Gamma(1 + \delta V) \frac{Z_2}{i \slashed{p}}, \quad
\label{eq:compgreenfunc} \langle(\overline{h}\Gamma
q)h\overline{q}\rangle_{\text{lat}} = \frac{Z^{\text{lat}}_h}{i
p_0}\Gamma(1 + \delta
V^{\text{lat}})\frac{(1-w^2_0)Z_w\,Z^{\text{lat}}_2}{i \slashed{p}},
\end{equation}

\noindent where the Feynman diagrams contributing at one-loop are
shown in Fig.~\ref{fig:heavylight}. All $Z$-factors have values $1 +
{\mathcal O}(\alpha_s)$, and the vertex corrections $\delta V,
\delta V^{\text{lat}}$ are ${\mathcal O}(\alpha_s)$ and
$\Gamma$-independent as noted above.

\paragraph{}

The continuum quantities are known \cite{Loktik:2006kz}; we focus on
the lattice correlation function: $w_0 = 1-M_5$ is a domain-wall
fermion specific constant, and an overlap factor $1-w^2_0$
connecting the five-dimensional and physical quark fields is present
even at tree level. The light quark wavefunction renormalization
$Z_w Z^{\text{lat}}_2$ due to Fig.~\ref{fig:heavylight} (a) and (b)
was calculated in Ref.~\cite{Aoki:2002iq}. $Z^{\text{lat}}_2$ can be
viewed as the four-dimensional wavefunction renormalization, while
$Z_w$ renormalizes the overlap factor $1-w^2_0$. Due to tadpoles,
the one-loop correction to $Z_w$ is enormous. As described in
Ref.~\cite{Aoki:2002iq}, this is remedied by reorganizing the
perturbation series according to the mean-field approach, resulting
in the prescriptions $M_5 \rightarrow \widetilde{M}_5 = M_5 -
4(1-u)$, $w_0 \rightarrow w_0^{\text{MF}} = 1 - \widetilde{M}_5$ and
$q^{\text{lat}} \rightarrow q^{\text{lat, MF}} = u^{-1/2}
q^{\text{lat}}$ to be made throughout the calculation; here $u =
P^{1/4}$ where $P$ is the measured average plaquette (for the
RBC-UKQCD calculation $u = 0.8757$) and the superscript \lq MF\rq~
identifies mean-field improved quantities. We calculate the matching
factor $\widetilde{C}_{A}(\mu,a^{-1})$ using

\begin{wrapfigure}{l}{5.5cm}
\begin{center}
\includegraphics{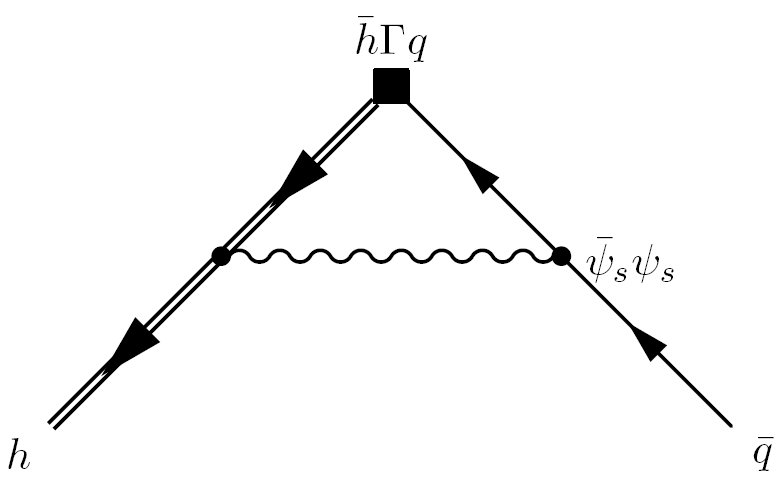}
\end{center}
\caption{\label{fig:heavylightvertex} One-Loop Vertex Correction to
the Heavy-Light Axial Current}
\end{wrapfigure}

\noindent both the usual continuum $\overline{\text{MS}}$ coupling
and a mean-field improved version, enabling an estimate of
${\mathcal O}(\alpha_s^2)$ corrections.
$\alpha_s^{\overline{\text{MS}}}(\mu)$ was obtained by running down
to the $c$ quark mass with the physical number of flavors and back
up to $\mu$ using only three dynamical flavors to match the
RBC-UKQCD $2+1$ flavor calculation:
$\alpha_s^{\overline{\text{MS}}}(\mu) = 0.326$ and
$\alpha_s^{\text{MF}}(\mu) = 0.177$. The calculation of the vertex
correction $\delta V^{\text{lat}}$ in Fig.~\ref{fig:heavylight} (c)
and the heavy quark wavefunction renormalization $Z_h^{\text{lat}}$
in Fig.~\ref{fig:heavylight} (d) and (e) is straightforward
\cite{AokiRBC:2007,Loktik:2006kz}. Infrared divergences only occur
in QED-like diagrams and are regulated by a gluon mass $\lambda$
which cancels from the matching factor. Furthermore, only the
unsmeared $\delta_{\mu0}$ part of $h_\mu(q)$ in
Eq.~(\ref{eq:formfactor}) gives rise to infrared divergences; the
sine functions in the smeared part of $h_\mu(q)$ cancel all infrared
divergent loop propagators. A generic feature of domain-wall fermion
perturbation theory is the appearance of correlation functions
$\langle q(-p) \overline{\psi}_s(p)\rangle$, $\langle
\psi_s(-p)\overline{q}(p)\rangle$ connecting external
four-dimensional quarks to five-dimensional quarks propagating in
loops, as shown in Fig.~\ref{fig:heavylightvertex}. A subtlety
pointed out in Ref.~\cite{Boucaud:1992nf} is that the correct
renormalization prescription for $Z^{\text{lat,\,MF}}_h$ includes
the linearly divergent heavy quark mass renormalization:



\begin{equation}
Z^{\text{lat,\,MF}}_h = 1 -i\frac{\partial \Sigma(p_0)}{\partial
p_0}\Big|_{p_0 = 0} + \Sigma(p_0 = 0),
\end{equation}

\noindent where the heavy quark self energy $\Sigma(p_0)$ itself is
not affected by mean-field improvement. Comparing the correlation
functions in Eq.~(\ref{eq:compgreenfunc}) after mean-field
improvement gives a matching factor

\begin{equation}
\widetilde{C}_{A}(\mu,a^{-1}) = \frac{\sqrt u}{\sqrt{(1-(w^{\rm
MF}_0)^2)Z^{\rm MF}_w}}\,
 Z^\mathrm{MF}_A(\mu,a^{-1}), \quad Z_A^{\text{MF}}(\mu,a^{-1}) = 1 +
 \frac{\alpha_s}{3\pi}(-1.584).
\end{equation}

\noindent The overall factor $Z_{\Phi}(\mu_b,a^{-1}) = C_{A}(\mu_b,
\mu)\widetilde{C}_{A}(\mu,a^{-1})$ relating the axial currents in
full QCD and the lattice static effective theory, computed using
both $\alpha_s^{\overline{\text{MS}}}(\mu)$ and
$\alpha_s^{\text{MF}}(\mu)$, is
$Z_{\Phi}^{\overline{\text{MS}}}(\mu_b,a^{-1}) = 0.902, \;
Z_{\Phi}^{\text{MF}}(\mu_b,a^{-1}) = 0.961$. While the one-loop
result is small and reliable, the large difference between
$\alpha_s^{\overline{\text{MS}}}(\mu)$ and
$\alpha_s^{\text{MF}}(\mu)$ induces a $\sim 7\%$ systematic error
ultimately warranting nonperturbative renormalization.

\paragraph{}

For completeness, we quote the lattice-continuum matching constants
(calculated in Refs.~\cite{AokiRBC:2007,Loktik:2006kz}) for the
four-fermion operators in Eqs.~(\ref{eq:statOVVAA}) and
(\ref{eq:statOSSPP}). For $i \in \{VV + AA, SS + PP\}$ and at
one-loop

\begin{equation}
\label{eq:fourfermionlattcont} \widetilde O_{i}(\mu) = \frac
u{(1-(w^{\rm MF}_0)^2)Z^{\rm MF}_w}\,
 Z^\mathrm{MF}_{i}(\mu,a^{-1}) a^{-6}O^{\rm lat}_{i}, \quad Z_{VV + AA}^{\text{MF}} = 1 +
 \frac{\alpha_s}{4\pi} (-4.462), \quad Z_{SS + PP}^{\text{MF}}
 = 1,
\end{equation}

\noindent where the $O_i^{\text{lat}}$ are dimensionless. Since the
coefficient $Z_2$ of $\widetilde O_{SS+PP}$ in
Eq.~(\ref{eq:full_vv+aa_matching}) is ${\mathcal O}(\alpha_s)$, only
the domain-wall overlap factors contribute to the lattice-continuum
matching for this operator. While formally inconsistent, we use the
one-loop mean-field improved values of the overlap factors
throughout to ensure tadpole-safety. Combining
Eqs.~(\ref{eq:full_vv+aa_matching}) and
(\ref{eq:fourfermionlattcont}) we get:

\begin{equation}
\label{eq:fullfourfermionmatching} O_{VV+AA} =
  Z_{VA}(\mu_b,a^{-1}) a^{-6}O^{\text{lat}}_{VV+AA} +
  Z_{SP}(\mu_b,a^{-1}) a^{-6}O^{\text{lat}}_{SS+PP},
\end{equation}
\begin{equation}
Z_{VA}^{\overline{\text{MS}}} = 0.902, \quad Z_{VA}^{\text{MF}} =
0.769, \quad Z_{SP}^{\overline{\text{MS}}} = -0.123, \quad
Z_{SP}^{\text{MF}} = -0.133.
\end{equation}

\section{Ground State Degeneracies of Static-Light Mesons and $f_B$, $B_B$ on the Lattice}

\paragraph{}

Let $H$ be the Hamiltonian corresponding to the full lattice action
in Sec.~\ref{sec:action}. For any $t$, the heavy quark action in
Eq.~(\ref{eq:hqa_imp}) is invariant under $h(\vec{x}) \rightarrow
e^{i\theta(\vec{x})}h(\vec{x})$ for a set of $V/a^3$ parameters
$\theta(\vec{x})$, where $V = L^3$ is the spatial lattice volume. If
$\Theta(\vec{x})$ is the generator corresponding to
$\theta(\vec{x})$ then

\begin{equation}
\label{eq:ThetaHComm} \left[\Theta(\vec{x}), h(\vec{y})\right] =
h(\vec{x}) \delta_{\vec{x}\vec{y}}, \quad \left[\Theta(\vec{x}),
\overline{h}(\vec{y})\right] = -\overline{h}(\vec{x})
\delta_{\vec{x}\vec{y}}, \quad
\left[\Theta(\vec{x}),\Theta(\vec{y})\right] = 0,\quad
\left[\Theta(\vec{x}), H\right] = 0.
\end{equation}

\noindent Simultaneously diagonalize $H$ and all $\Theta(\vec{x})$.
Since Eq.~(\ref{eq:ThetaHComm}) implies that $h(\vec{x})$ and
$\overline{h}(\vec{x})$ raise and lower the eigenvalues of
$\Theta(\vec{x})$ by $1$, and the charge conjugation invariance of
QCD implies $\Theta(\vec{x})|0\rangle = 0$, the spectrum of
$\Theta(\vec{x})$ contains $\mathbb{Z}$. Define the unit-norm state
$|B(\vec{x})\rangle$ to be the lowest energy state with the quantum
numbers of a $B$ meson which also satisfies
$\Theta(\vec{y})|B(\vec{x})\rangle =
\delta_{\vec{x}\vec{y}}|B(\vec{x})\rangle$. Thus $\langle B(\vec{x})
| B(\vec{y}) \rangle = \delta_{\vec{x}\vec{y}}$, and we can
interpret these states as having the heavy quark localized at a
fixed lattice site with the light quark smeared out around it. Since
$T(\hat{i}\,) \Theta(\vec{x}) T(\hat{i}\,)^{-1} = \Theta(\vec{x} +
\hat{i}\,)$, where $T(\hat{i}\,)$ is a lattice translation by $a$ in
the spatial direction $\hat{i}$, all $B$ meson ground states
$|B(\vec{x})\rangle$ are degenerate. We also define total spatial
momentum eigenstates $|\widetilde{B}(\vec{k}_l)\rangle$, where $l_i
\in \mathbb{Z} \;\; (i = 1,2,3)$:

\begin{equation}
|\widetilde{B}(\vec{k}_l)\rangle = \sqrt{2a^3} \sum_{\vec{x}}
e^{-i\,\vec{k}_l\cdot\vec{x}}|B(\vec{x})\rangle, \;\; \vec{k}_l =
\frac{2\pi}{L}(l_1,l_2,l_3), \;\; -\frac{L}{2a} < l_i \leq
\frac{L}{2a}, \;\; \langle \widetilde{B}(\vec{k}_{l'}) |
\widetilde{B}(\vec{k}_{l})\rangle = 2V\delta_{l'l}.
\end{equation}

\noindent As $a\rightarrow 0, V \rightarrow \infty$, these states
reduce to continuum momentum eigenstates
$|\widetilde{B}(\vec{p})\rangle^{\text{c}}$ with conventional static
effective theory normalization
$^{\text{c}}\langle\widetilde{B}(\vec{p}\,')|\widetilde{B}(\vec{p})\rangle^{\text{c}}
= 2 (2\pi)^3 \delta^{(3)}(\vec{p}\,' - \vec{p})$. In the $m_b
\rightarrow \infty$ limit, these states only differ from the
corresponding full QCD states by a factor of $\sqrt{m_B}$. Thus:

\begin{align}
f_B\sqrt{m_B} \equiv \langle 0 |
A_{0}(\vec{0},0)|\widetilde{B}(\vec{p} = \vec{0})\rangle^{\text{c}}
&= Z_{\Phi}^{\text{MF}}a^{-3}\langle 0 |
A_{0}^{\text{lat}}(\vec{0},0)\left(\sqrt{2a^3} \sum_{\vec{x}}
|B(\vec{x})\rangle \right) = \nonumber \\ &=
\sqrt{2}Z_{\Phi}^{\text{MF}}a^{-3/2}\langle
0|A_0^{\text{lat}}(\vec{0},0)|B(\vec{0})\rangle \equiv
\sqrt{2}Z_{\Phi}^{\text{MF}}a^{-3/2}\Phi_B^{\text{lat}}.
\quad\quad\;\;\;
\end{align}

\noindent In complete analogy to the above, we can construct
$\overline{B}$ meson ground states states
$|\overline{B}(\vec{x})\rangle$. Using these and
Eq.~(\ref{eq:fullfourfermionmatching}), the calculation of the $B -
\overline{B}$ mixing matrix element $\frac{8}{3}\,m^2_B \,f_B^{\,2}
\,B_B = \langle \, \overline{B}|O_{VV+AA}|B\rangle$ is reduced to
the calculation of the lattice quantities $\langle
\overline{B}(\vec{0})
|O^{\text{lat}}_{i}(\vec{0},0)|B(\vec{0})\rangle, \; i \in \{VV +
AA, SS + PP\}$.

\paragraph{}

The degeneracy of the states $|B(\vec{x})\rangle$ complicates the
extraction of $\Phi_B^{\text{lat}}$ and $\langle\,
\overline{B}(\vec{0})
|O^{\text{lat}}_{i}(\vec{0})|B(\vec{0})\rangle$, since even a large
time separation of source and sink may not project onto a unique $B$
meson ground state: different combinations of the
$|B(\vec{x})\rangle$ may enter the correlation functions used for
calculating the matrix elements and those used for normalization. To
see this, consider the extraction of $\Phi_B^{\text{lat}}$; we now
work exclusively in the lattice theory. Define local and smeared $B$
meson interpolation operators $A^{L}_{0}(\vec x, t) =
\overline{h}(\vec x, t)\gamma_{0}\gamma_{5}q(\vec x, t) $,
$A^{S}_{0}(t) = \sum_{\vec y \in \Delta V} \sum_{\vec z \in \Delta
V}
        \overline{h}(\vec y,t)\gamma_0 \gamma_5 q(\vec z,t)$, where
$\Delta V$ is a fixed subvolume of $V$ and the smeared operators are
Coulomb gauge fixed. From experience, local-local correlation
functions in the static effective theory are prohibitively noisy;
instead calculate the local-smeared and smeared-smeared correlation
functions. Inserting a complete set of states $\sum_{\vec{w}}
|B(\vec{w})\rangle\langle B(\vec{w})| + (\text{higher energy
states})$ with the correct quantum numbers, we have as $t
\rightarrow \infty$:

\begin{equation}
\label{eq:LScorrelator} \mathcal C^{LS}(t) \equiv \sum_{\vec x \in
V}
  \langle0|A^L_0(\vec x,t)A^S_0(0)^\dagger|0\rangle = \Phi_B^{\rm{lat}} e^{-m_B^{*}\,t}
\left(\sum_{\vec{w} \in V} \langle B(\vec{w})|\sum_{\vec{y} \in
\Delta V}\sum_{\vec{z} \in \Delta V}\overline{q}(\vec{y},0) \gamma_0
\gamma_5 h(\vec{z},0)|0\rangle\right),
\end{equation}

\begin{equation}
\mathcal C^{SS}(t) \equiv
 \langle0|A^S_0(t)A^S_0(0)^{\dagger}|0\rangle = e^{-m_B^{*}\,t} \left( \sum_{\vec{w} \in
V}\Big|\langle B(\vec{w})|\sum_{\vec{y} \in \Delta V}\sum_{\vec{z}
\in \Delta V}\overline{q}(\vec{y},0) \gamma_0 \gamma_5
h(\vec{z},0)|0\rangle\Big|^2\right).
\end{equation}

\noindent where $m_B^{*}$ is the unphysical mass of the lattice $B$
meson. Since ${\mathcal C}^{SS}(t)$ contains a sum over squares, the
use of a naive ratio $\sim {\mathcal C}^{LS}(t)/\sqrt{{\mathcal
C}^{SS}(t)}$ requires a translationally invariant wall source
$\Delta V = V$ to project onto the unique state of zero-momentum. In
this case the sums over $\vec{w}$ only give a factor of $V/a^3$ and
$\Phi_B^{\text{lat}} = {\mathcal C}^{LS}(t)/\sqrt{{\mathcal
C}^{SS}(t)e^{-m_B^{*}\,t}\,V/a^3}$. To remedy the poor overlap of
the wall source with the $B$ meson ground state - especially on
large lattices - consider a fixed box source and a series of box
sinks summed over an entire timeslice to project onto zero momentum;
this approach also allows more general types of smearing, such as
the use of an atomic wavefunction. Let
$\widetilde{A}^{S}_{0}(\vec{w}, t) = \sum_{\vec y \in \Delta
V_{\vec{w}}} \sum_{\vec z \in \Delta V_{\vec{w}}}
        \overline{h}(\vec y,t)\gamma_0 \gamma_5 q(\vec z,t)$ where
        $\Delta V_{\vec{w}}$ is a box of fixed size located at
        $\vec{w}$ and $\Delta V_{\vec{0}} = \Delta V$, $\widetilde{A}^{S}_{0}(\vec{0}, t) = A^S_0(t)$. Define a corresponding smeared-smeared correlation
        function and insert a complete set of momentum eigenstates $\frac{1}{2V} \sum_{\vec{k}_l}
|\widetilde{B}(\vec{k}_l)\rangle \langle \widetilde{B}(\vec{k}_l)| +
(\text{higher energy states})$; then as $t \rightarrow \infty$,

\begin{equation}
\mathcal C^{\widetilde{S}\widetilde{S}}(t) \equiv
 \sum_{\vec{w}}\langle0|\widetilde{A}^S_0(\vec{w},t)A^S_0(0)^{\dagger}|0\rangle = \frac{e^{-m_B^{*}\, t}}{2V}\sum_{\vec{w}}\langle 0 |
 \widetilde{A}^S_0(\vec{w},t) | \widetilde{B}(\vec{0}) \rangle \langle
 \widetilde{B}(\vec{0}) | A^S_0(0)^{\dagger}|0\rangle =
 \frac{e^{-m_B^{*} \, t}}{2a^3}\Big | \langle
 \widetilde{B}(\vec{0}) | A^S_0(0)^{\dagger}|0\rangle \Big |^2.
\end{equation}

\noindent Since $| \widetilde{B}( \vec{0} ) \rangle = (2 a^3)^{1/2}
\sum_{\vec{w}} | B ( \vec{ w } ) \rangle$, we can rewrite the right
side of Eq.~(\ref{eq:LScorrelator}) and obtain another ratio for
$\Phi_B^{\rm{lat}}$ which reaches a plateau more quickly due to the
improved ground state overlap:

\begin{equation}
\label{eq:boxphibcalc} \mathcal C^{\,LS}(t)e^{m_B^{*} \, t/2}\Big
/{\sqrt{\mathcal C^{\widetilde{S}\widetilde{S}}(t)}} =
 \Phi_B^{\rm{lat}}\langle
 \widetilde{B}(\vec{0}) | \,A^S_0(0)^{\dagger}|0\rangle\Big / \sqrt{\Big | \langle
 \widetilde{B}(\vec{0}) | A^S_0(0)^{\dagger}|0\rangle \Big |^2} =
 \Phi_B^{\rm{lat}}.
\end{equation}

\noindent The calculation of $\langle \overline{B}(\vec{0})
|O^{\text{lat}}_{i}(\vec{0})|B(\vec{0})\rangle$, $i \in \{VV + AA,
SS + PP\}$ is considerably simpler. Define

\begin{equation}
\mathcal C_{O_i}(T,t) \equiv  \sum_{\vec x \in V}\langle 0| \,
\overline{A}^{\,S}_{0}(T)O^{\, \rm lat}_{i}(\vec x,t) \,
A^{S}_{0}(0)^{\dagger}|0\rangle,
\end{equation}

\noindent where $\overline{A}^{\,S}_{0}(T) = \sum_{\vec y \in \Delta
V} \sum_{\vec z \in \Delta V}
        \overline{q}(\vec y,T)\gamma_0 \gamma_5 h(\vec z,T)$. Proceeding as above, we have as $\,t, \; T - t \rightarrow \infty$:

\begin{equation}
\label{eq:boxmixcalc} \langle \, \overline{B}(\vec{0}) |
O^{\text{lat}}_i(\vec{0},0)|B(\vec{0})\rangle = \mathcal
C_{O_i}(T,t)\Big / \mathcal C^{SS}(T) = \mathcal C_{O_i}(T,t)
e^{m_B^{*}T/2} \Big / \sqrt{\mathcal C^{SS}(T - t)\mathcal
C^{SS}(t)}.
\end{equation}

\noindent Here no zero momentum projection is necessary; the use of
$\mathcal C^{SS}$ for smaller time separations simply reduces noise.
Using Eqs.~({\ref{eq:boxphibcalc}) and (\ref{eq:boxmixcalc}) we can
thus calculate $f_B$ and $B_B$ using only box sources and sinks.
These are preferable to wall sources, whose poor ground state
overlap led to late plateaus in the $V = (2 \; \rm{fm})^3$ RBC-UKQCD
calculation and presents an even bigger problem for the ongoing
extension to $V =(3 \; \rm{fm})^3$. It is worth emphasizing that
this simple method relies on the particular properties of the static
effective theory, and further such improvements might be possible.

\acknowledgments

We thank our RBC-UKQCD collaborators C. Albertus, Y. Aoki, P. A.
Boyle, L. Del Debbio, J. M. Flynn, C. T. Sachrajda, A. Soni, and J.
Wennekers. We gratefully acknowledge the support of BNL, Columbia
University, the University of Edinburgh, PPARC, RIKEN, and the U.S.
DOE.

\bibliography{cdil_lat07}

\providecommand{\href}[2]{#2}\begingroup\raggedright\begin{thebibliography}{1}

\bibitem{Wennekers:2007}
C.~Albertus, Y.~Aoki, P.~A. Boyle, N.~H. Christ, L.~Del~Debbio, T.~T.
  Dumitrescu, J.~M. Flynn, T.~Izubuchi, O.~Loktik, C.~T. Sachrajda, A.~Soni and
  J.~Wennekers, {\it {$B-\bar{B}$ mixing with domain wall fermions}},  {\em
  {$\rm{in \; these \; proceedings, \;}$}} {\bf {\pos{PoS(LATTICE 2007)376}}}.

\bibitem{AokiRBC:2007}
C.~Albertus, Y.~Aoki, P.~A. Boyle, N.~H. Christ, L.~Del~Debbio, T.~T.
  Dumitrescu, J.~M. Flynn, T.~Izubuchi, O.~Loktik, C.~T. Sachrajda, A.~Soni and
  J.~Wennekers, {\it {Unquenched $B$ meson decay constants and $B^0 -
  \bar{B}^0$ mixing parameters from chiral latttice QCD}},  {\em {$\rm{in} \;
  \rm{preparation}$}}.

\bibitem{Aoki:2007}
{Y. Aoki for the RBC and UKQCD collaborations}, {\it Heavy-light matrix
  elements in static limit with domain wall fermions},  {\em {$\rm{in \; these
  \; proceedings, \;}$}} {\bf {\pos{PoS(LATTICE 2007)345}}}.

\bibitem{Loktik:2006kz}
O.~Loktik and T.~Izubuchi, {\it Perturbative renormalization for static and
  domain-wall bilinears and four-fermion operators with improved gauge
  actions},  {\em Phys. Rev.} {\bf D75} (2007) 034504
  [\href{http://arXiv.org/abs/hep-lat/0612022}{{\tt hep-lat/0612022}}].

\bibitem{Aoki:2002iq}
S.~Aoki, T.~Izubuchi, Y.~Kuramashi and Y.~Taniguchi, {\it {Perturbative
  renormalization factors in domain-wall QCD with improved gauge actions}},
  {\em Phys. Rev.} {\bf D67} (2003) 094502
  [\href{http://arXiv.org/abs/hep-lat/0206013}{{\tt hep-lat/0206013}}].

\bibitem{Boucaud:1992nf}
P.~Boucaud, J.~P. Leroy, J.~Micheli, O.~Pene and G.~C. Rossi, {\it {A rigorous
  treatment of the lattice renormalization problem of $f_B$}},  {\em Phys.
  Rev.} {\bf D47} (1993) 1206--1218
  [\href{http://arXiv.org/abs/hep-lat/9208004}{{\tt hep-lat/9208004}}].

\end{thebibliography}\endgroup

\end{document}